\documentclass[aps,prc,twocolumn,superscriptaddress,showpacs,floatfix,nofootinbib]{revtex4-1}
\usepackage{amsmath,graphicx}
\usepackage{color}

\usepackage{ulem}

\begin{document}
\title{Azimuthal Anisotropy Distributions in High-Energy Collisions}

\author{Li Yan}
\author{Jean-Yves Ollitrault}
\affiliation{
CNRS, URA2306, IPhT, Institut de physique th\'eorique de Saclay, F-91191
Gif-sur-Yvette, France} 
\author{Arthur M. Poskanzer}
\affiliation{Lawrence Berkeley National Laboratory, Berkeley,
California, 94720}
\date{\today}

\begin{abstract}
Elliptic flow in ultrarelativistic heavy-ion collisions results 
from the hydrodynamic response 
to the spatial anisotropy of the initial density profile. 
A long-standing problem in the interpretation of flow data is 
that uncertainties in the initial anisotropy are mingled with
uncertainties in the response. 
We argue that the non-Gaussianity of flow fluctuations in small systems
with large fluctuations can be used to disentangle the initial state from the response. We apply this method to recent measurements of 
anisotropic flow in Pb+Pb and p+Pb collisions at the LHC, 
assuming linear response to the initial anisotropy. 
The response coefficient is found to decrease 
as the 
system becomes smaller and is consistent with a low value of the
ratio of viscosity over entropy of $\eta/s\simeq 0.19$.
Deviations from linear response are studied. 
While they significantly change the value of the response coefficient
they do not change the rate of decrease with centrality. Thus,
we argue that the estimate of $\eta/s$ is robust against
non-linear effects. 
\end{abstract}

\maketitle

\section{Introduction}
\label{sec:intro}

The large magnitude of elliptic flow, $v_2$, in relativistic heavy-ion
collisions at RHIC~\cite{Ackermann:2000tr} and LHC~\cite{Aamodt:2010pa}
has long been recognized as a signature of hydrodynamic behavior of
the strongly-interacting quark-gluon plasma~\cite{Kolb:2003dz}. 
$v_2$ is understood as the hydrodynamic response to the 
initial anisotropy, $\varepsilon_2$, of the initial density profile~\cite{Ollitrault:1992bk}.
However, the magnitude of this anisotropy is poorly constrained
theoretically~\cite{Hirano:2005xf,Lappi:2006xc}. This uncertainty
hinders the extraction of the properties of the quark-gluon
plasma from experimental data~\cite{Luzum:2008cw,Heinz:2013th}. 

The statistical properties of anisotropic flow are now precisely
known~\cite{Jia:2014jca}. The ATLAS collaboration has analyzed the full
probability distribution of $v_2$, $v_3$ and $v_4$ in Pb+Pb collisions
for several centrality windows~\cite{Aad:2013xma}. 
In p+Pb collisions, information is less detailed, but 
the first moments of the distribution of $v_2$ have been
measured~\cite{Chatrchyan:2013nka}.   
Our goal is to make use of these measurements 
to separate the initial state from the
response  without assuming any particular model of the initial conditions -- by only using a simple functional form which goes to zero at the geometric limits of $\varepsilon_n=0$ and $1$.

In theory, one can describe the particles emitted from a collision with an underlying probability distribution~\cite{Luzum:2013yya}. 
Anisotropic flow, $v_n$, is defined as the $n^{\rm th}$ Fourier coefficient of the azimuthal probability distribution ${\cal P}(\varphi)$:
\begin{equation}
\label{defVn}
V_n=v_n  e^{in\Psi_n}\equiv \frac{1}{2\pi}\int_0^{2\pi} {\cal P}(\varphi)e^{in\varphi}d\varphi,
\end{equation}
where we have used a complex notation~\cite{Ollitrault:2012cm,Teaney:2013dta}. 
Note that the underlying probability distribution ${\cal P}(\varphi)$ and $V_n$ fluctuate event to event, 
but they are both theoretical quantities which cannot be measured on an event-by-event basis. 
The particles that are detected in an event represent a finite sample of ${\cal P}(\varphi)$, and 
the measurement of the probability distribution of $v_n$ involves a nontrivial unfolding of statistical fluctuations~\cite{Aad:2013xma}.

\section{Distribution of $\varepsilon_n$}
\label{sec:EP}

We assume that the fluctuations of $v_n$  for $n=2,3$ are due to
fluctuations of the initial anisotropy $\varepsilon_n$ in the
corresponding harmonic, defined by~\cite{Teaney:2010vd} 
\begin{equation}
\label{defepsilon}
{\mathcal E}_n=
\varepsilon_n e^{in\Phi_n}
\equiv -\frac{\int r^n e^{in\phi}\rho(r,\phi)r{\rm d}r{\rm d}\phi}
{\int r^n\rho(r,\phi)r{\rm d}r{\rm d}\phi}.
\end{equation}
where $\rho(r,\phi)$ is the
energy density near midrapidity shortly after the collision, and  
$(r,\phi)$ are polar coordinates in the transverse plane, in a
coordinate system where the energy distribution is centered at the origin.

We assume for the moment that $v_n$ in a given event is determined 
by linear response to the initial anisotropy, $v_n=\kappa_n\varepsilon_n$, 
where $\kappa_n$ is a response coefficient which does not fluctuate event to event. Event-by-event hydrodynamic calculations~\cite{Niemi:2012aj} show that
this is a very good approximation for $n=2,3$.
Within this approximation, it has already been shown
that one can rule out particular models of the initial 
density using either a combined
analysis~\cite{Alver:2010dn,Retinskaya:2013gca} 
of elliptic flow and triangular flow~\cite{Alver:2010gr} data, 
or the relative magnitude of elliptic flow 
fluctuations~\cite{Alver:2006wh,Alver:2010rt,Renk:2014jja}. 
Our goal is to show that one can
extract both $\kappa_n$ and the distribution of
$\varepsilon_n$ from data. We hope to show that this is true even if we relax the linear assumption.
We make use of the recent observation that the distribution 
of $\varepsilon_n$ is to a large extent 
universal~\cite{Yan:2013laa,Yan:2014afa} and can be characterized by 
two parameters. 

\begin{figure*}
\includegraphics[width=.9\linewidth]{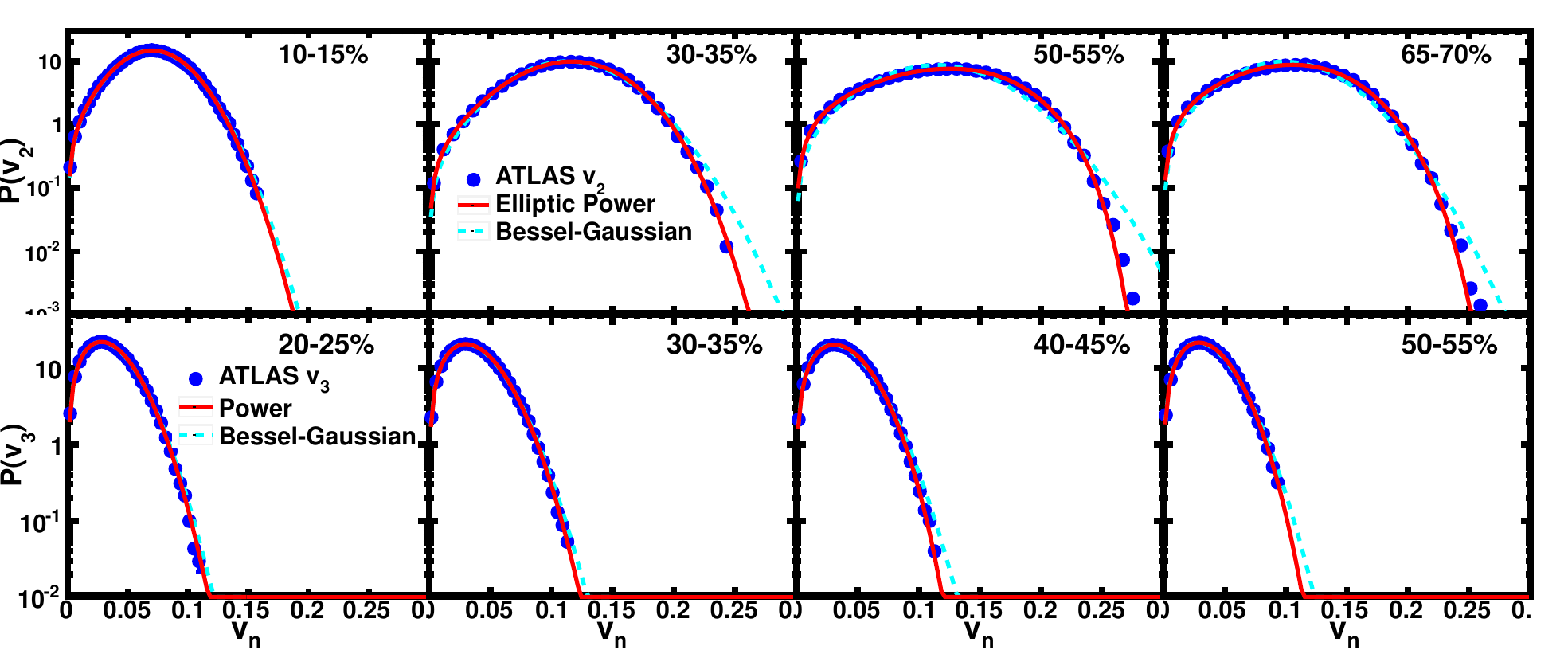}
\caption{ (Color online) Distribution of $v_2$ (top) and $v_3$ (bottom) in various centrality windows. 
Symbols: ATLAS data~\cite{Aad:2013xma} for Pb+Pb collisions at $\sqrt{s_{NN}}=2.76$~TeV. 
For $v_2$, fits are rescaled Elliptic Power Eq.~(\ref{ellipticpower}) (full lines) and Bessel-Gaussian distributions Eq.~(\ref{besselgaussian}) (dashed lines). 
For $v_3$, fits are rescaled Power Eq.~(\ref{power}) (full lines) and 
Bessel-Gaussian distributions with $\varepsilon_0=0$ Eq.~(\ref{vngaussian})  (dashed lines). 
}
\label{fig:atlasfits}
\end{figure*}

Both the magnitude and direction of ${\mathcal E}_n$ fluctuate event to event. 
The simplest parametrization of these fluctuations is a two-dimensional 
Gaussian probability distribution which, upon integration over azimuthal angle, yields the Bessel-Gaussian distribution~\cite{Voloshin:2007pc}:
\begin{equation}
\label{besselgaussian}
p(\varepsilon_n)=\frac{\varepsilon_n}{\sigma^2}I_0\left(\frac{\varepsilon_0\varepsilon_n}{\sigma^2}\right)
\exp\left(-\frac{\varepsilon_0^2+\varepsilon_n^2}{2\sigma^2}\right),
\end{equation}
where $\varepsilon_0$ is the mean anisotropy in the reaction plane, which vanishes by 
symmetry for odd $n$, and $\sigma$ is the typical magnitude of eccentricity fluctuations around this mean anisotropy. 
Both $\varepsilon_0$ and $\sigma$ depend on the harmonic $n$. 

In a previous publication~\cite{Yan:2014afa}, we have introduced an alternative parametrization, the Elliptic Power distribution: 
\begin{equation}
\label{ellipticpower}
p(\varepsilon_n)=
\frac{2\alpha\varepsilon_n}{\pi}
  (1-\varepsilon_0^2)^{\alpha+\frac{1}{2}}
  \int_0^{\pi}\frac{(1-\varepsilon_n^2)^{\alpha-1} d\phi}
{(1-\varepsilon_0\varepsilon_n\cos\phi)^{2\alpha+1}},
\end{equation}
where $\alpha$ describes the fluctuations and is approximately proportional to the number of 
sources in an independent-source model~\cite{Ollitrault:1992bk}.
The parameter $\alpha$ depends on $n$. 
When $\varepsilon_0\ll 1$ and $\alpha\gg 1$, Eq.~(\ref{ellipticpower}) reduces to Eq.~(\ref{besselgaussian})  with $\sigma\approx 1/\sqrt{2\alpha}$. 
Its support is the unit disk: it naturally takes into account the condition $|\varepsilon_n|\le 1$ 
which follows from the definition, Eq.~(\ref{defepsilon}). 
For this reason, it is a better parametrization than the Bessel-Gaussian, in particular for large anisotropies. 
Eq.~(\ref{ellipticpower}) has been shown to fit various initial-state models~\cite{Yan:2014afa}.
Note that $\varepsilon_0$ is not strictly equal to the mean reaction plane eccentricity for the Elliptic Power 
distribution, but the difference is small for Pb+Pb collisions~\cite{Yan:2014afa}. 

When the anisotropy is solely due to fluctuations, $\varepsilon_0=0$, 
the Bessel-Gaussian reduces to a Gaussian distribution: 
\begin{equation}
\label{vngaussian}
p(\varepsilon_n)=\frac{\varepsilon_n}{\sigma^2}
\exp\left(-\frac{\varepsilon_n^2}{2\sigma^2}\right),
\end{equation}
and the Elliptic Power distribution 
reduces to the Power distribution~\cite{Yan:2013laa}:
\begin{equation}
\label{power}
p(\varepsilon_n)=
2\alpha\varepsilon_n
(1-\varepsilon_n^2)^{\alpha-1}. 
\end{equation}

\section{Distribution of $v_n$}
\label{sec:vn}
The probability distribution of anisotropic flow, $P(v_n)$, is obtained from the distribution of the initial anisotropy $p(\varepsilon_n)$ by
\begin{equation}
\label{probav}
P(v_n)=\frac{d \varepsilon_n}{d v_n}p\!\left(\varepsilon_n\right).
\end{equation}
Assuming $v_n=\kappa_n\varepsilon_n$, this becomes:
\begin{equation}
\label{probavn}
P(v_n)=\frac{1}{\kappa_n}p\!\left(\frac{v_n}{\kappa_n}\right).
\end{equation} 
In this case the distribution is rescaled by the response coefficient $\kappa_n$.
Figure~\ref{fig:atlasfits} displays the probability distribution of $v_2$ and $v_3$ in various centrality windows~\cite{Aad:2013xma} together with fits using rescaled Bessel-Gaussian and Elliptic Power distributions for $v_2$, and rescaled Gaussian and Power distributions for $v_3$. 
Both parametrizations give very good fits to $v_2$ and $v_3$ data for the most central bins shown on the figure.\footnote{The fits do not converge below $10\%$  ($20\%$) centrality for $v_2$ ($v_3$), which reflects the fact that the distributions become very close to Bessel-Gaussian (Gaussian).}
As the centrality percentile increases, however, the quality of the Bessel-Gaussian fit becomes 
increasingly worse, which is reflected by the large $\chi^2$ of the fit, and also clearly seen in the 
tail of the distribution: it systematically overestimates the distribution for 
large anisotropies. 
On the other hand, the Elliptic Power fit is excellent for all centralities. 
In particular, it falls off more steeply for large $v_n$, in close agreement with the data. 

Note that the Bessel-Gaussian distribution Eq.~(\ref{besselgaussian}) is scale invariant: rescaling it by $\kappa_n$  amounts to multiplying both $\varepsilon_0$ and $\sigma$ by $\kappa_n$, so that the fit is degenerate: only the products $\kappa_n\varepsilon_0$ and $\kappa_n\sigma$ can be determined. Therefore the Bessel-Gaussian fit to ATLAS data is in practice a 2-parameter fit for $v_2$, and a 1-parameter fit for $v_3$. 
On the other hand, the Elliptic Power fit is not degenerate because of the non-Gaussian cut-off at $\varepsilon_n=1$, and returns both the response coefficient 
$\kappa_n$ and the parameters pertaining to the shape, namely $\alpha$ and $\varepsilon_0$ (for $v_2$). 
However, the fit parameters are still correlated in the sense that the combinations $\kappa_n/\sqrt{\alpha}$ and  $\kappa_n\varepsilon_0$ (for $v_2$)  have much smaller errors than each individual parameter.

\section{Experimental Errors}
\label{sec:errors}
Our ability to separate the response from the initial eccentricity thus lies in the difference between the Bessel-Gaussian and the Elliptic Power fits, that is, in the non-Gaussianity of flow fluctuations. 
Since the difference is small, errors must be carefully evaluated. 

The ATLAS collaboration reports the statistical error, the systematic error on the mean $\langle v_n\rangle$, and the systematic error on the relative standard deviation $\sigma_{v_n}/\langle v_n\rangle$.
The first systematic error is an error on the scale of the distribution, while the second is an error on its shape. 
The error on the scale directly translates into an error of the response coefficient $\kappa_n$, of the same relative magnitude. 
Since our analysis uses the deviations from a Gaussian shape, the dominant source of error is ---by far--- the error on the shape.  
In order to estimate the corresponding error on our fit parameters, we
distort the distribution of $v_n$ in such a way that the mean $\langle
v_n\rangle$ is unchanged, and $\sigma_{v_n}/\langle v_n\rangle$ is
increased or decreased by the experimental uncertainty. This is done in practice by shifting the values of the $v_n$ bins according to  
$v_n\to v_n+\delta(v_n)$, where $\delta(v_n)$ is a small non-linear shift. We choose the ansatz $\delta(v_n)=\epsilon v_n(v_{\rm max}-v_n)(v_n-\lambda)$, where $v_{\rm max}$ is the tail of the $v_n$ distribution, $\lambda$ is chosen in such a way that 
$\langle v_n\rangle$ is unchanged, and $\epsilon$ is chosen in such a way that $\sigma_{v_n}/\langle v_n\rangle$ is increased or decreased by the systematic error.
This non-linear transformation leaves the minimum and maximum values of $v_n$ invariant.

\begin{figure}
\includegraphics[width=\linewidth]{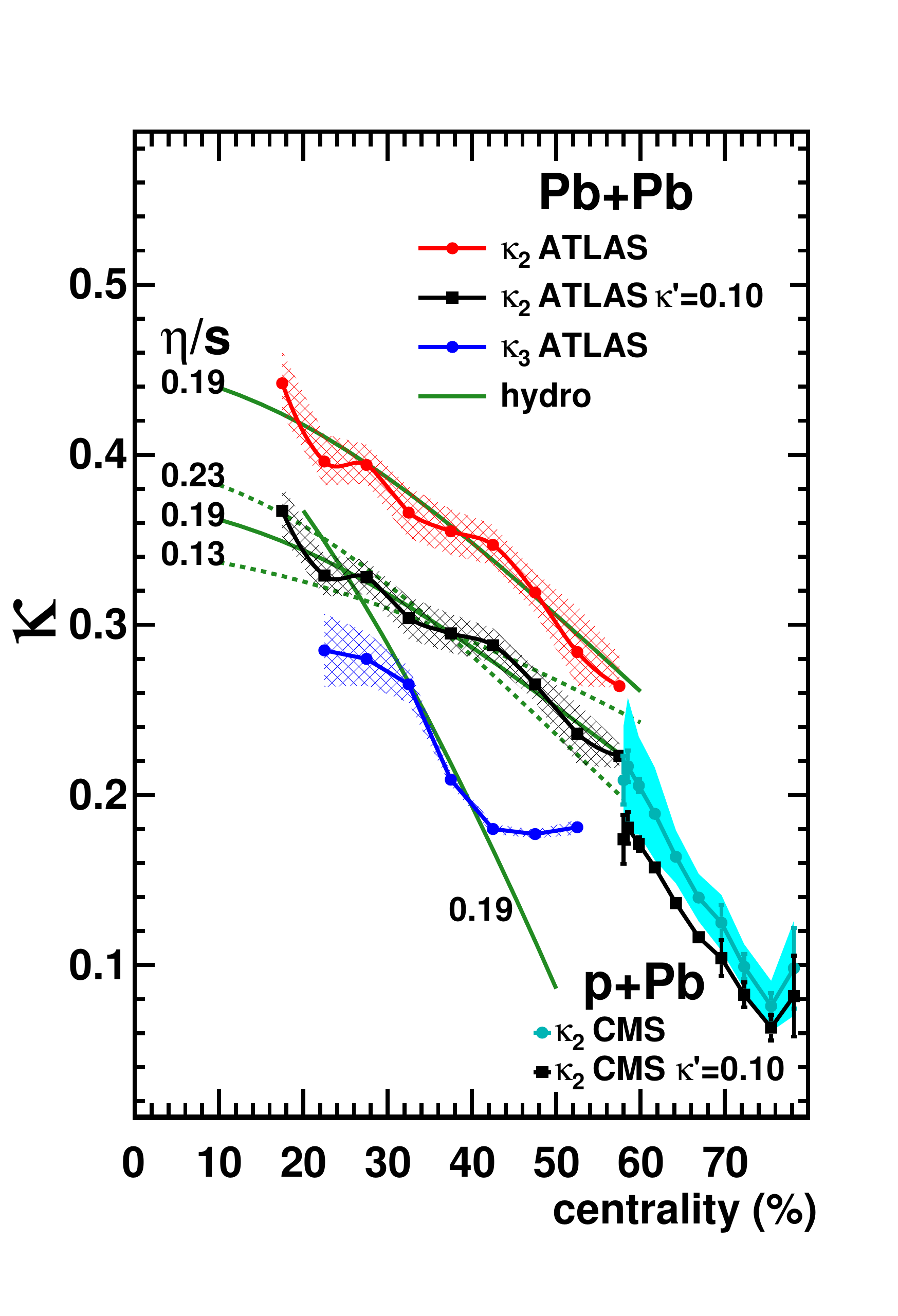}
\caption{ (Color online) Response coefficients $\kappa_2$ and $\kappa_3$  versus centrality.
Symbols:  results from the fits to ATLAS Pb+Pb data~\cite{Aad:2013xma} and to CMS p+Pb data~\cite{Chatrchyan:2013nka}. 
For $\kappa_2$, systematic and statistical experimental errors are
added in quadrature. For $\kappa_3$, only the statistical error is
shown. Also shown are $\kappa_2$ values from fitting the $v_2$
distributions with a non-linear term caracterized in Eqs.~(\ref{cubic}
and \ref{powerCubic}) by $\kappa'$. The smooth solid lines are the result of a viscous hydrodynamic calculation for $\kappa$ with
$\eta/s=0.19$. The upper solid line is normalized up by the factor $1.7$,
the middle line by the factor $1.4$, and the lower line ($\kappa_3$) by the factor $3.2$.  
The dashed lines for $\kappa_2$ with $\kappa'=0.1$ are shown for comparison: they are for hydro results with $\eta/s=0.13$ (normalized up by $1.2$) and $\eta/s=0.23$ (normalized up by $1.6$). 
}
\label{fig:kappa}
\end{figure}

\section{Parameters of the Distributions}
\label{sec:results}
Figure~\ref{fig:kappa} displays the value of the response coefficients $\kappa_2$ and $\kappa_3$ as a function of the centrality percentile. 
They are smaller than unity, with $\kappa_3<\kappa_2$, in line with
expectations from hydrodynamic calculations~\cite{Niemi:2012aj}, and
decrease 
as a function of the centrality percentile, which is the general behavior expected from viscous corrections to local equilibrium~\cite{Voloshin:1999gs,Drescher:2007cd}. 
We estimate that the low-$p_T$ cut of ATLAS at 0.5~GeV increases $\kappa_2$ by a factor 1.4 to 1.5. 
The systematic error for $\kappa_3$ is very large and therefore not shown: for most bins, the upper error bar goes all the way to infinity. 
Now, if one takes the limit $\kappa_3\to\infty$ while keeping the rms $v_3$ constant, $\alpha$ in Eq.~(\ref{power}) also goes to infinity and the Power distribution reduces to a Gaussian distribution Eq.~(\ref{vngaussian}). Therefore  
the ATLAS $v_3$ distributions are compatible with Gaussians within systematic errors.

\begin{figure*}
\includegraphics[width=.48\linewidth]{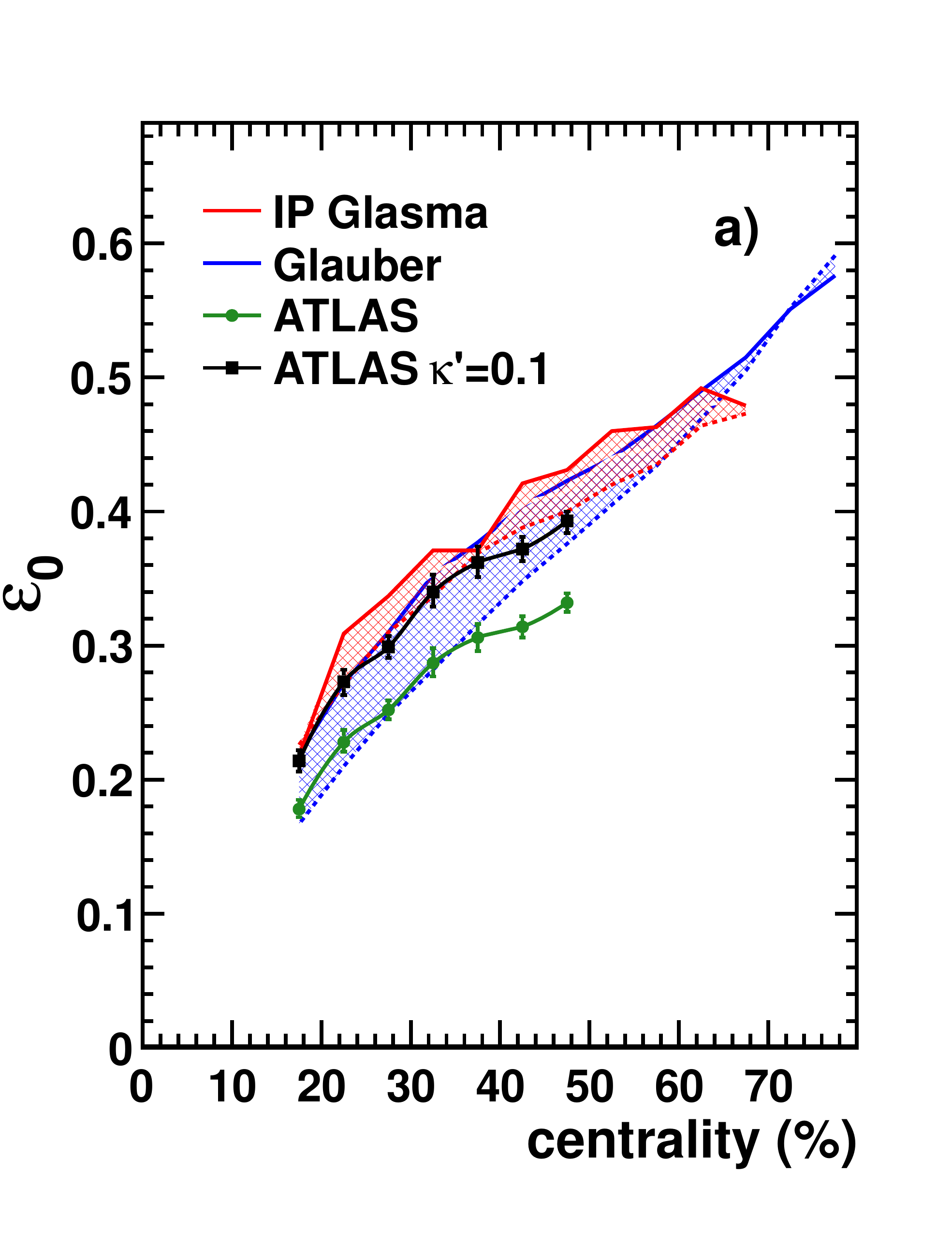}
\includegraphics[width=.48\linewidth]{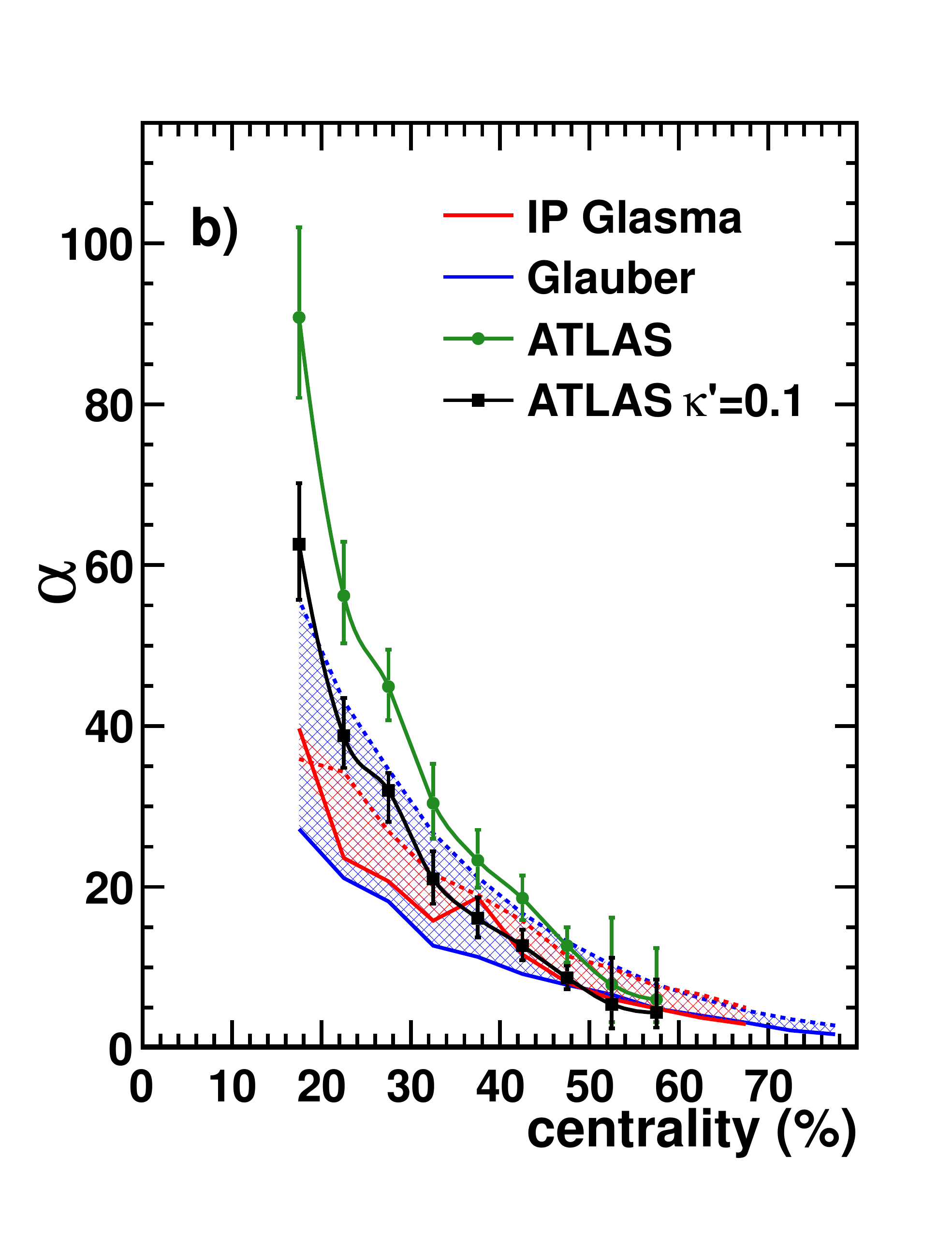}
\caption{ (Color online)  $\varepsilon_0$ (a) and $\alpha$ (b) 
versus centrality. 
Symbols: results from the fits to ATLAS $v_2$ data.  
Predictions from the Monte-Carlo Glauber~\cite{Alver:2008aq} and 
IP-Glasma~\cite{Schenke:2012fw,Schenke:2013aza} models are shown as shaded bands. Also shown are values from fits with a non-linear term defined by $\kappa'$ in Eq.~(\ref{cubic}).
}
\label{fig:eps0alpha}
\end{figure*}

The other parameters of the fit to $v_n$ distributions, namely,
$\varepsilon_0$ and $\alpha$, 
characterize the shape of the distribution.
They are displayed in Fig.~\ref{fig:eps0alpha} 
as a function of the collision centrality. 
$\varepsilon_0$ increases smoothly with the centrality percentile: extrapolation to the most central collisions 
(where the fit does not converge) gives $\varepsilon_0=0$, as required by azimuthal symmetry. 

Figure~\ref{fig:eps0alpha} also displays comparisons
with the Glauber~\cite{Miller:2007ri,Broniowski:2007ft,Alver:2008aq}
and IP-Glasma~\cite{Schenke:2012fw,Schenke:2013aza} models. These models are shown as shaded bands.  
The bands correspond to the fact that the Elliptic Power distribution does not exactly fit 
the distribution of $\varepsilon_2$ for that particular model. 
Specifically, the dashed line at the edge of the band is the value returned by a 2-parameter Elliptic Power fit to the distribution of $\varepsilon_2$. 
The full line at the other edge of the band is the value 
that the fit to the $v_2$ distribution would return if $v_2\propto\varepsilon_2$, with $\varepsilon_2$ given by that model. 
If one assumes linear response, ATLAS data deviate from both models.

\section{Cumulants}
\label{sec:cum}
For p+Pb collisions, the full distribution of $v_2$ has not been measured, but only its first 
cumulants~\cite{Borghini:2001vi,Bilandzic:2010jr} $v_2\{2\}$ and $v_2\{4\}$~\cite{Aad:2013fja,Chatrchyan:2013nka,Abelev:2014mda}. 
Assuming linear response to the initial eccentricity, each measured cumulant is proportional to the corresponding cumulant of the initial eccentricity~\cite{Miller:2003kd},
$v_2\{k\}=\kappa_2\varepsilon_2\{k\}$, for  $k=2,4,6$...
The eccentricity in p+Pb collisions is solely due to fluctuations~\cite{Bozek:2011if,Bozek:2012gr}, 
therefore Eqs.~(\ref{vngaussian}) and (\ref{power}) apply. 
While cumulants of order 4 and higher vanish for the Gaussian distribution Eq.~(\ref{vngaussian}), 
the Power distribution Eq.~(\ref{power}) always gives $\varepsilon_n\{4\}>0$~\cite{Yan:2013laa}. 
We again use this non-Gaussianity to disentangle the initial state from the response: We extract $\alpha$ from the measured ratio
 $v_n\{4\}/v_n\{2\}\simeq\varepsilon_n\{4\}/\varepsilon_n\{2\}=(1+\frac{\alpha}{2})^{-1/4}$~\cite{Yan:2013laa}. 
 The rms anisotropy is then obtained as  
$\varepsilon_n\{2\}=1/\sqrt{1+\alpha}$~\cite{Yan:2013laa}. One finally obtains for the Power distribution: 
\begin{equation}
\label{kappa_pPb}
\kappa_n=\frac{v_n\{2\}}{\varepsilon_n\{2\}}=
v_n\{2\}\sqrt{2\left(\frac{v_n\{2\}}{v_n\{4\}}\right)^4-1}.
\end{equation} 
The values of $\kappa_2$ extracted from CMS p+Pb data~\cite{Chatrchyan:2013nka} using this equation are also displayed in Fig.~\ref{fig:kappa}.
We multiply them by a factor $1.19$ to correct for the different low-$p_T$ cut ($0.3$~GeV/$c$) assuming a linear dependence of $v_2$ on $p_T$.
We plot p+Pb data 
at the equivalent centralities, determined according to the number of charged tracks~\cite{Chatrchyan:2013nka}. 
General arguments have been put forward which suggest that the hydrodynamic response should be identical for p+Pb and Pb+Pb at the same equivalent centrality~\cite{Basar:2013hea}. 
Once rescaled, the p+Pb slope is in line with 
Pb+Pb results, albeit somewhat steeper. 

Note that the fit parameters can also be obtained from cumulants for  
$v_3$ in Pb+Pb collisions  using Eq.~(\ref{kappa_pPb}). 
For $v_2$, there is a third parameter $\varepsilon_0$,  therefore one needs a third cumulant $v_2\{6\}$. 
$\alpha$ and $\varepsilon_0$, which control the shape of the distribution and its non-Gaussian features,
can be extracted from the ratios $v_2\{6\}/v_2\{4\}$ and $v_2\{4\}/v_2\{2\}$ using the Elliptic Power distribution (Eq.~(A5) of Ref.~\cite{Yan:2014afa}). Note that while the Bessel-Gaussian Eq.~(\ref{besselgaussian}) gives $\varepsilon_n\{4\}=\varepsilon_n\{6\}=\varepsilon_0$~\cite{Bzdak:2013rya}, the Elliptic Power distribution always gives $\varepsilon_n\{6\}<\varepsilon_n\{4\}$. We have checked that $\alpha$ and $\varepsilon_0$ thus extracted from cumulant ratios are essentially identical to those obtained by fitting the distribution of $v_2$.
This approach has the advantage that  cumulants can be analyzed without any unfolding procedure~\cite{Bilandzic:2010jr} but $v_2\{2\}$ may suffer from non-flow effects.

\section{Deviations from linear scaling}
\label{sec:non-lin}
We now discuss the effect of deviations from linear eccentricity scaling of anisotropic flow. 
Because of such deviations, the shape of the $v_n$ distribution is not exactly the same as  that of the 
$\varepsilon_n$ distribution, as already noted in event-by-event hydrodynamic calculations~\cite{Schenke:2014zha}.
There are two distinct types of non-linearities: $v_n$ can be a function of $\varepsilon_n$ which is not exactly linear, or $v_n$ can depend on properties of the initial state other than $\varepsilon_n$. We study these effects in turn. 

Adding a quadratic term would be equivalent to rotating the distribution 90 degrees or changing the sign of $v_n$. Thus the first significant non-linear term is the cubic:
\begin{equation}
\label{cubic}
	v_{2} = \kappa_{2} \varepsilon_{2} + \kappa'\kappa_{2}\varepsilon_{2}^3 .
\end{equation}
Several hydrodynamic calculations show evidence that $\kappa'>0$~\cite{Bhalerao:2005mm,Alver:2010dn,Paatelainen:2014}, but no quantitative analysis has been done yet. 
One typically expects $\kappa'$ to depend mildly on centrality. 
When fitting the experimental $v_2$ distributions with the added parameter of the 
cubic term, $\kappa'$ had large errors but was in the range from 0 to 0.15. Thus we fixed $\kappa'$ at 
0.10 and plotted the $\kappa_2$ values also in Fig.~\ref{fig:kappa}. 
The effect of the non-linear response is to reduce the linear response coefficient $\kappa_2$, essentially by a constant factor. 
In the case where the distribution of  $\varepsilon_{2}$ is the Power distribution (\ref{power}) and in the limit $\alpha\gg 1$, 
the relative change of $\kappa_2$ is
\begin{equation}
\label{powerCubic}
	\frac{\Delta \kappa_2}{\kappa_2} = -2\kappa' 
\end{equation}
to leading order.\footnote{This result is obtained by inserting Eq.~(\ref{cubic}) into Eq.~(\ref{kappa_pPb}) and using the approximate relation $\varepsilon_2\{4\}^4\simeq \langle\varepsilon_2^6\rangle-2 \langle\varepsilon_2^4\rangle\langle\varepsilon_2^2\rangle$ for $\alpha\gg 1$.}
With the Elliptic-Power distribution, the relative effect is also $\sim -2\kappa'$ as can be
seen in Fig.~\ref{fig:kappa}. 
Note that this non-linear correction to the response is much larger than one would naively expect from Eq.~(\ref{cubic}):
the relative magnitude of the cubic term $\kappa'\varepsilon_2^2\ll\kappa'$, yet it produces an effect of order $\kappa'$. 
The reason is that the non-linear response contributes to the non-Gaussianity of flow fluctuations. 
 
We now discuss deviations from linearity due to the fact that $v_n$ is not entirely determined by $\varepsilon_n$~\cite{Teaney:2010vd}.
One can generally decompose the flow as  
$V_n=\kappa_n{\mathcal E}_n+X_n$, 
where $X_n$ is uncorrelated with the initial eccentricity ${\mathcal E}_n$.
There can be various contributions to $X_n$ from non-linear coupling between different harmonics~\cite{Teaney:2012ke} or radial modulations of the initial density~\cite{Teaney:2010vd}. 
In order to estimate their effect on the hydrodynamic response, we further assume that $X_n$ is a Gaussian noise. 
Then, the distribution of $V_n$ is a rescaled Elliptic-Power distribution, convoluted with a Gaussian: the deviation from linearity here results in a Gaussian smearing of the distribution.

A quantitative measure of the magnitude of $X_n$ is the Pearson correlation coefficient $r_n$ between 
the anisotropic flow and the initial anisotropy, defined as 
\begin{eqnarray}
\label{pearson}
r_n&\equiv&\frac{\langle V_n{\mathcal E}_n^*\rangle}{\sqrt{\langle |V_n|^2\rangle\langle |{\mathcal E}_n|^2\rangle}}\cr
&=&\left(1+\frac{\langle |X_n|^2\rangle}{\kappa_n^2\langle |{\mathcal E}_n|^2\rangle}\right)^{-1/2},
\end{eqnarray}
where angular brackets denote an average value over events in a centrality class.
Our analysis assumes the maximum correlation, $|r_n|=1$.
Event-by-event hydrodynamic calculations show that there are small deviations around eccentricity scaling~\cite{Holopainen:2010gz}. 

Ideal hydrodynamics~\cite{Gardim:2011xv} gives $|r_2|\sim 0.95$ for
elliptic flow. 
However, the correlation between $v_n$ and $\varepsilon_n$ has been shown to be significantly larger
in viscous hydrodynamics~\cite{Niemi:2012aj}, and a value $|r_2|=0.99$ seems reasonable, 
but there is to date no quantitative estimate of $|r_2|$ as defined
in  Eq.~(\ref{pearson}).
 
The effect on the fit parameters can be obtained using the fact that the rms flow $v_n\{2\}=\sqrt{\langle |V_n|^2\rangle}$ 
is increased by the noise, while higher-order cumulants $v_n\{4\}$ and $v_n\{6\}$ (see below Sec.~\ref{sec:cum}) are unchanged. 
We find that a decrease of  $|r_2|$ by 1\% results in an decrease of
the extracted $\kappa_2$ by 6\% to 9\%, depending on the centrality,
the effect being maximum in the 20-30\% centrality range. 
The value of $|r_2|$ found in ideal hydrodynamic
calculations~\cite{Gardim:2011xv} depends mildly on centrality and is
closest to 1 also in the 20-30\% centrality range. 
Therefore one can conjecture ---this should eventually be confirmed by
detailed calculations--- that the effect of the noise $X_n$ is to
reduce the extracted response essentially by a constant factor,
independent of centrality. 

Note that the cubic response in Eq.~(\ref{cubic}) does not contribute to $r_2$ to first order in $\kappa'$, 
so that the two effects are in practice well separated. 

The conclusion is that  deviations from linear eccentricity scaling
all make $\kappa_2$ smaller, by a factor which can be significant, but
depends little on centrality. This is of crucial importance for the
extraction of the viscosity over entropy ratio (see below). 
The decrease in $\kappa_2$ makes $\varepsilon_0$ larger 
and $\alpha$ smaller (see Fig.~\ref{fig:eps0alpha}), thereby improving
compatibility with existing initial-state models.

\section{Viscous Hydro}
\label{sec:hydro}
We now compare our result for $\kappa_2$ with hydrodynamic
calculations. To the extent that anisotropic flow scales linearly with
eccentricity, the value of the response coefficient $\kappa_2$ is
independent of initial conditions. 
In ideal hydrodynamics, scale invariance implies that $\kappa_2$ is independent 
of the system size, i.e., independent of centrality. 
Deviations from thermal equilibrium generally result in a reduction of the flow which is stronger for peripheral collisions~\cite{Voloshin:1999gs,Drescher:2007cd}.
In a hydrodynamic calculation~\cite{Heinz:2013th}, such deviations are due to the shear 
viscosity~\cite{Luzum:2008cw} 
and, to a lesser extent, to the freeze-out procedure at the end of the hydrodynamic expansion. 
Therefore the dependence of $\kappa_2$ on centrality in Fig.~\ref{fig:kappa} can be used to estimate the 
shear viscosity over entropy ratio $\eta/s$ of the quark-gluon plasma. 
We use the same hydrodynamic code as in Ref.~\cite{Teaney:2012ke} to estimate $\kappa_2$. The resulting values 
are significantly smaller than the data in Fig.~\ref{fig:kappa}. 
Since we have shown that deviations from linear eccentricity scaling reduce $\kappa_2$ without altering its centrality dependence, we compensate for this effect, and the low $p_T$ cut of the ATLAS data, by multiplying our hydrodynamic result by a constant, while tuning the viscosity so as to match the centrality dependence of $\kappa_2$.  
Since the systematic errors on $\kappa_3$ are so large, we only fit
$\kappa_2$. 
The smooth solid lines in Fig.~\ref{fig:kappa} are
obtained with $\eta/s=0.19$. 
The dashed lines show the sensitivity to $\eta/s$. This extracted value of $\eta/s=0.19$ is consistent with that reported in the literature~\cite{Heinz:2013th}, using specific models of the initial state.
For sake of illustration, we also show the result for $\kappa_3$ with
the same $\eta/s$ and the same overall normalization factor as for
$\kappa_2$. 
We recall that systematic errors on $\kappa_3$ from experimental data
are very large so that no conclusion on $\eta/s$ can be drawn from these data
alone. 

\section{Summary}
\label{sec:sum}
We have shown that a rescaled Elliptic Power distribution fits  
the measured distributions of elliptic and triangular flows in Pb+Pb collisions at the LHC. 
These distributions become increasingly non-Gaussian as the anisotropy increases.  
We have used this non-Gaussianity to disentangle for the first time 
the initial anisotropy from the response without assuming any particular model of initial conditions -- just using a simple functional form which meets the geometrical constrants of eccentricity.  
 
This is another aspect of the analogy between heavy-ion physics and 
cosmology~\cite{Mishra:2007tw,Dusling:2011rz}, where initial quantum fluctuations give 
rise to correlations, and the non-Gaussian statistics of these correlations can be used to unravel the properties of the initial state~\cite{Maldacena:2002vr,Bartolo:2004if,Ade:2013ydc}.
The non-Gaussianity is stronger for smaller systems, which is an incentive to analyze flow in smaller collision systems. 

We have found that the hydrodynamic response to ellipticity 
has the expected overall magnitude and centrality dependence: 
it decreases 
with centrality percentage. 
A somewhat similar slope is found for p+Pb collisions. 
This decrease can be attributed to the viscous suppression of $v_2$. 
Comparison with hydrodynamic calculations supports a low value of the
viscosity over entropy ratio, $\eta/s\sim 0.19$.

The present study can be improved by constraining the cubic response 
coefficient $\kappa'$ in Eq.~(\ref{cubic}) as well as the Pearson
coefficient due to  other non-linear terms in Eq.~(\ref{pearson}). 
This could be done in future hydrodynamic calculations. 
Taking into account these nonlinear terms will decrease the magnitude
of the response and therefore improve the agreement with hydrodynamic
calculations.  
However, we have argued that this decrease is essentially a constant
factor, independent of centrality, so that our estimate of $\eta/s$ is
likely to be robust. 
Our study is a first step toward the extraction of the viscosity over
entropy ratio of the quark-gluon plasma from experimental data,
without any prior knowledge of the initial state.

\begin{acknowledgments}
We thank M. Luzum and S. Voloshin for extensive discussions and suggestions. 
In particular, we thank S. Voloshin for useful comments on the manuscript.
JYO thanks the MIT LNS for hospitality. 
LY is funded  by the European Research Council under the 
Advanced Investigator Grant ERC-AD-267258. AMP was supported by the Director, Office of Nuclear Science of the U.S. Department of Energy.
\end{acknowledgments}

\end{document}